\documentclass[10pt, prd,aps,twocolumn,a4paper,floatfix,showpacs, nofootinbib]{revtex4-1}

\usepackage[utf8]{inputenc}  
\usepackage[T1]{fontenc}
\usepackage{graphicx,psfrag}
\usepackage{mathrsfs}
\usepackage{amsmath,amsfonts,amssymb,pifont,gensymb}
\usepackage{multirow,enumerate}
\usepackage{comment,hyperref}
\usepackage{color}
\usepackage{acronym}
\usepackage{xspace}
\usepackage[normalem]{ulem}
\usepackage{mathtools}
\usepackage{subfigure}
\usepackage{makecell}
\usepackage{appendix}

\usepackage{color}
\definecolor{cyan}{rgb}{0,0.9,0.9}
\definecolor{orange}{rgb}{0.9,0.5,0}
\definecolor{magenta}{rgb}{1,0,1}
\definecolor{purple}{rgb}{0.8,0.4,0.8}
\definecolor{gray}{rgb}{0.5,0.5,0.5}
\definecolor{mygreen}{rgb}{0.1,0.8,0.1}
\definecolor{darkblue}{rgb}{0.0,0.0,0.6}

\newcommand{\boldtheta}{\boldsymbol{\theta}}
\newcommand{\boldphi}{\boldsymbol{\phi}}

\begin{document}

\title{Pre-Merger Detection and Characterization of Inspiraling Binary Neutron Stars Derived from Neural Posterior Estimation}
\author{Wouter van Straalen$^{1}$}
\author{Alex Kolmus$^{2}$}
\author{Justin Janquart$^{1,3,4,5}$}
\author{Chris Van Den Broeck$^{1,3}$}

\affiliation{${}^1$ Department of Physics, Utrecht University, Princetonplein 1, 3584 CC Utrecht, The Netherlands}
\affiliation{${}^2$ Institute for Computing and Information Sciences (ICIS), Radboud University Nijmegen, Toernooiveld 212, 6525 EC Nijmegen, The Netherlands}
\affiliation{${}^3$ Nikhef, Science Park 105, 1098 XG Amsterdam, The Netherlands}
\affiliation{${}^4$  Center for Cosmology, Particle Physics and Phenomenology - CP3, Universit\'e Catholique de Louvain, Louvain-La-Neuve, B-1348, Belgium}
\affiliation{${}^5$ Royal Observatory of Belgium, Avenue Circulaire, 3, 1180 Uccle, Belgium}

\date{\today}

\begin{abstract}
    \noindent
    As the sensitivity of the international gravitational wave detector network increases, observing binary neutron star signals will become more common. Moreover, since these signals will be louder, the chances of detecting them before their mergers increase. However, this requires an efficient framework. In this work, we present a machine-learning-based framework capable of detecting and analyzing binary neutron star mergers during their inspiral. Using a residual network to summarize the strain data, we use its output as input to a classifier giving the probability of having a signal in the data, and to a normalizing-flow network to perform neural posterior estimation. We train a network for several maximum frequencies reached by the signal to improve the estimate over time. Our framework shows good results both for detection and characterization, with improved parameter estimation as we get closer to the merger time. Thus, we can effectively evolve the precision of the sky location as the merger draws closer. Such a setup would be important for future multi-messenger searches where one would like to have the most precise information possible, as early as possible. 
\end{abstract}

\maketitle

\section{Introduction}

On the 17$^{\mathrm{th}}$ of August 2017, the first binary neutron star (BNS) gravitational wave (GW) signal (GW170817)~\cite{LIGOScientific:2017vwq} was detected by the LIGO~\cite{LIGO} and Virgo~\cite{VIRGO} detectors. About two seconds later, a gamma-ray burst~\cite{LIGOScientific:2017zic} was observed by the Fermi Gamma-Ray Burst Monitor~\cite{Meegan_2009} and the INTEGRAL satellite~\cite{2017ApJ...848L..15S}. However, this was found in an archival search since the LIGO-Virgo collaboration published an estimated sky location only several hours after detection. Based on this, multiple telescopes performed electromagnetic (EM) follow-up studies. This detection opened up the new field of multi-messenger astronomy (MMA), which combines observations of high-energy neutrinos, ultra-high-energy cosmic rays, gamma rays, EM channels, and GW data~\cite{Meszaros:2019xej}. Among other things, GW170817 provided us with an independent measurement of the Hubble constant~\cite{LIGOScientific:2017adf}, constraints on the neutron star equation of state~\cite{LIGOScientific:2018cki}, and a comparison between the speed of light and gravity~\cite{LIGOScientific:2017zic}. 

While GW170817 already provided a plethora of discoveries, we could do better by observing other phases of the signal in the EM band. If we can detect and locate BNS signals before the merger, we could observe the EM emission during the inspiral and merger phases. This can be useful in several ways; for example, to study the pre-merger magnetospheric interactions between neutron stars~\cite{Most:2020ami}, further investigation of the r-process, which is the production mechanism for the heavy elements in the Universe~\cite{2017ApJ...848L..18N}, and examining X-ray emissions at the merger to determine its final state~\cite{Metzger:2013cha}.

With upcoming detector upgrades~\cite{collaborationreport}, KAGRA~\cite{KAGRA:2013rdx} joining the global network, and the construction of next-generation detectors like Cosmic Explorer~\cite{Reitze:2019iox} and Einstein Telescope~\cite{Punturo:2010zz, Sathyaprakash:2012jk}, the detector network's sensitivity will increase significantly. Consequently, BNS signals will be detected more often and will be louder~\cite{KAGRA:2013rdx, Maggiore:2019uih, Branchesi:2023mws}. Moreover, for the detectors with a lower frequency reach, the BNS signals would also stay in band for longer, ranging from tens of minutes for A$^{\#}$-like detectors~\cite{collaborationreport}, to hours for third-generationd detectors~\cite{Reitze:2019iox, Punturo:2010zz, Sathyaprakash:2012jk}. These combined effects will strongly increase the opportunities for pre-merger sky localization in the future.

Several studies have demonstrated the ability to provide early alerts for BNS inspirals using the LIGO and Virgo detectors; some rely on more classical matched filtering-based methods~\cite{Sachdev:2020lfd, Chu:2020pjv, Adams:2015ulm} while others use convolutional neural networks to produce triggers~\cite{Baltus:2021nme, Yu:2021vvm, Baltus:2022pep}. These studies can provide early alert triggers up to several minutes before the merger, allowing for follow-up investigations to estimate the sky location before the merger. The latter requires a follow-up with a(n approximate) parameter estimation framework.

The most significant hurdle in pre-merger sky localization is ensuring that all necessary steps are completed swiftly. Conventional parameter estimation methods typically rely on Bayesian inference methods such as Markov Chain Monte-Carlo (MCMC)~\cite{Veitch:2014wba} and nested sampling~\cite{2020MNRAS.493.3132S, Romero-Shaw:2020owr}. While these methods converge to the optimal posteriors given enough time, their computational cost makes them unsuitable for pre-merger sky localization. There are alternative methods capable of producing sky maps rapidly. First, there is \textsc{bayestar}~\cite{Singer:2015ema}, which relies on matched-filtering outputs and uses the Fisher matrix formalism to produce sky maps within seconds. This is nominally used for complete signals but can also be adapted to produce sky maps when only the early inspiral of the signal is observed~\cite{Sachdev:2020lfd}. The second framework is \textsc{GWSkyLocator}~\cite{Chatterjee:2022dik}, a machine-learning (ML) method that maps the time series of the signal-to-noise ratio (SNR) of the best-matched template and its corresponding intrinsic parameters to a sky map. While efficient, these methods rely on the output of matched filtering pipelines to generate the sky map. However, the template bank used is discrete and the best matching template can be off in real noise. Additionally, these methods only provide a point estimate value of the intrinsic parameters, not accounting for the potentially large uncertainty in the early phase of the signal. Finally, extra latency originates from the need to combine multiple pipelines.

In recent years, ML-based methods have gained increased prominence in GW science, seeing use in glitch classification, glitch cancellation, denoising, and binary black hole (BBH) signal detection~\cite{Cuoco:2020ogp, Benedetto:2023jwn}. In parameter estimation, multiple studies have used NF-based neural networks~\cite{2019arXiv191202762P, Dax:2021tsq, Dax:2022pxd, Langendorff:2022fzq, Gupte:2024jfe, Kolmus:2024scm, Wouters:2024oxj} to do neural posterior estimation (NPE) due to their speed compared to classical methods while arguably being equally accurate~\cite{Dax:2022pxd}. Besides traditional parameter estimation~\cite{2019arXiv191202762P, Dax:2021tsq, Dax:2022pxd}, they have been used to tackle the problem of overlapping signals~\cite{Langendorff:2022fzq}, approximating proposal distribution in nested sampling~\cite{Williams:2021qyt}, or enhancing MCMC~\cite{Gabrie:2021tlu, Wong:2022xvh}.

In this work, we propose an ML framework capable of detecting and localizing inspiraling BNS mergers without the need for matched filtering. We start by summarizing the strain data with a context network. Its output is then passed to a multi-layer perception (MLP) to determine whether a signal is present in the data or not. If a signal is detected, the output of the context network is passed to an NF-based network to perform parameter estimation, which provides estimations for the sky map, the masses, the luminosity distance, and the inclination. These parameters are chosen as they are the most important to determine whether an EM counterpart could be detected\footnote{In the future, additional parameters can easily be detected.}. So, our system is completely independent of other traditional approaches reducing the latency related to inter-pipeline communication. Moreover, since it is ML-based, the bulk of the computational cost is moved up front and detection and characterization can be done in about a second.

This paper is organized as follows: in Sec.~\ref{sec:bns_theory} we introduce the main quantities used to characterize the signals in this work. The ML algorithms and the setup used to do detection and inference are detailed in Sec.~\ref{sec:analysis_setup}. We detail the results for detection, inference, and the entire framework in Secs.~\ref{sec:detection_results}, \ref{sec:npe_results}, and \ref{sec:joint_results}, respectively. Finally, in Sec.~\ref{sec:concl}, we summarize our findings and suggest future work directions. 

\section{Characteristics of Binary Neutron Star Signals}
\label{sec:bns_theory}

\begin{figure}
    \centering
    \includegraphics[width = 0.5 \textwidth]{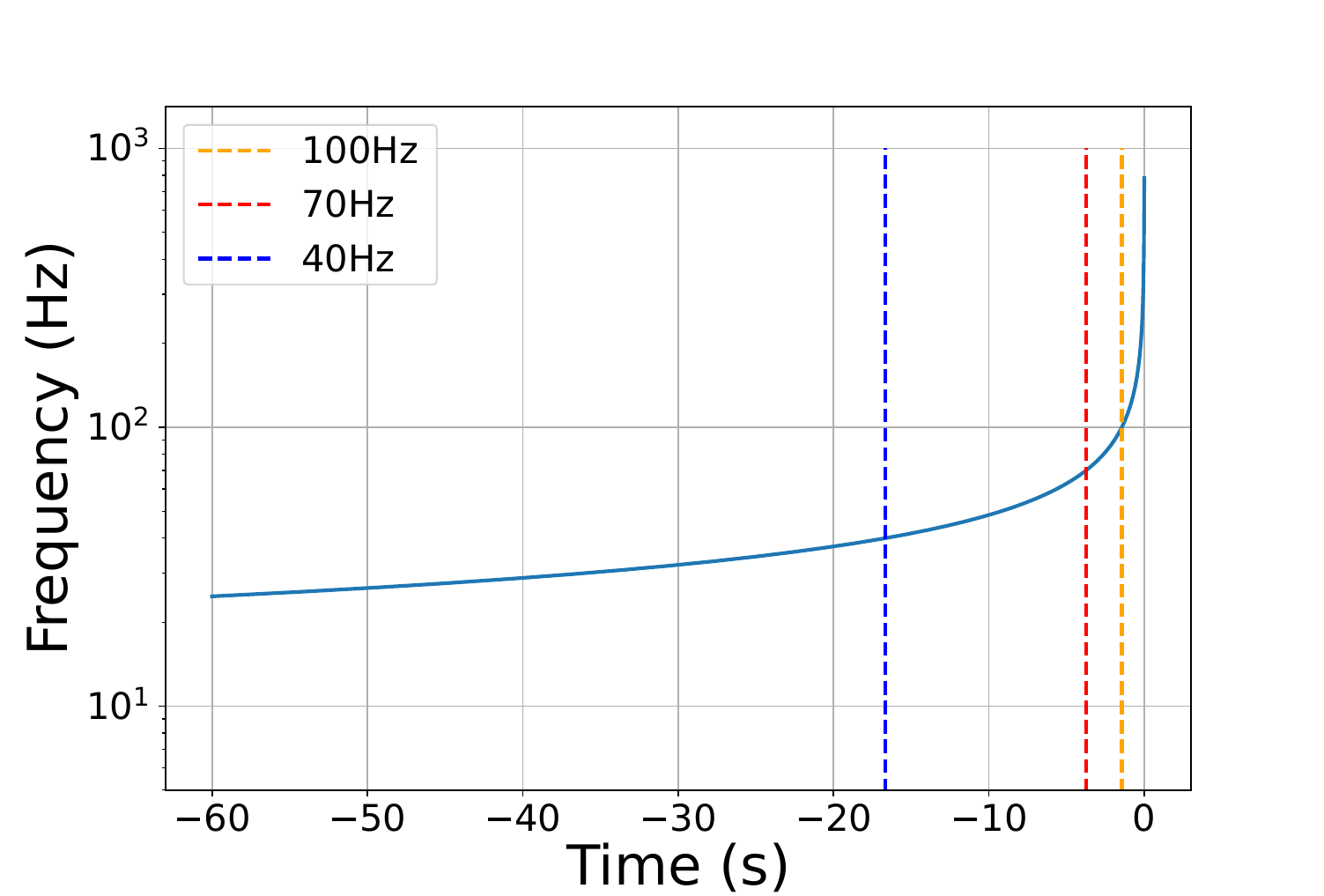}
    \caption{Frequency evolution of a CBC signal with $\mathcal{M}_{c} = 1.8\,M_{\odot}$ as described in Eq.~\eqref{eq:freq_evolution}, corresponding to the evolution at lowest frequency. $t = 0$ corresponds to the time at which the approximation breaks down. }
    \label{fig:freq_evolution}
\end{figure}

For pre-merger detection and characterization, we focus on BNS signals since those are the ones we most expect to produce an EM counterpart\footnote{Neutron star black hole mergers could also give an EM counterpart, depending on the model, but tend to be shorter, leaving less time for pre-merger detection. Nevertheless, the method developed in this work is easily adaptable to such signals.}. Such signals, with low component masses, stay in band for a long time, from several minutes in current detectors to hours in next-generation ones. At the lowest order in velocity, the time taken by the signal to go from a frequency $f_{min}$ to a frequency $f_{max}$ is:
\begin{equation}\label{eq:time_freq_evolution}
    \tau \simeq 2.18 \bigg(\frac{1.21\,M_{\odot}}{\mathcal{M}_c} \bigg) \bigg[ \bigg(\frac{100 \, \mathrm{Hz}}{f_{min}}\bigg)^{8/3} -  \bigg(\frac{100 \, \mathrm{Hz}}{f_{max}}\bigg)^{8/3} \bigg] \, \mathrm{s},
\end{equation}
where $\mathcal{M}_c$ is the chirp mass of the  signal. Since BNSs have a low chirp mass, they take more time to go from one frequency to another. Moreover, inverting this equation, one can also obtain the frequency evolution with time:
\begin{equation}
    f(t) = \frac{1}{\pi} \left( \frac{G\mathcal{M}_{c}}{c^{2}}  \right)^{-5/8} \left( \frac{5}{251} \frac{1}{t_{c} - t} \right) \, \mathrm{Hz} ,
    \label{eq:freq_evolution}
\end{equation}
where $t_c$ is the time of coalescence, roughly corresponding to the time when the two signals merge; again, this shows that low masses evolve slower but also that the time required to go from a frequency $f$ to a frequency $f + \Delta f$ is larger for lower frequencies, translating the ``chirping'' nature of CBC signals. This non-linear time and frequency behavior is illustrated in Fig.~\ref{fig:freq_evolution}.

Here, we work with data in the frequency domain, i.e. we give the sky maps and event characteristics for a given maximum signal frequency. As can be seen from Eqs.~\eqref{eq:time_freq_evolution} and~\eqref{eq:freq_evolution}, the time left before the merger will be different depending on the mass of the system, with more time left for lower mass signals. We chose to represent the data in the frequency domain because it enables us to have a fixed data array to pass to our neural network. Indeed, when collecting data, the strain is in the time domain and we take a Fourier transform before doing the analysis. Therefore, if the signal is in band up to a certain time, it is the same as having it up to a certain frequency in this other representation and only noise at higher frequencies. So, the frequency representation enables us to keep a fixed frequency representation while accommodating all the cases we want to analyze, solving the potential issue of data with varying sizes. This could ease the process of moving to the use of a single network dealing with all possible maximum frequencies, though this is left for future work.

Another important quantity for training neural networks is the loudness of the signal since louder signals are easier to detect and characterize. Generally, the loudness of a signal is described by its SNR ($\rho$), defined as 
\begin{equation}
    \rho = \left(4 \Re\left(\int_{f_{\text{min}}}^{f_{\text{max}}} \frac{\tilde{d}(f)\tilde{h}^{*}(f)}{S_{n}(f)}  \right)\right)^{1/2}\,,
    \label{eq:SNR}
\end{equation}
where $d$ is the data, $h$ the waveform for which we do the calculation, tilde means the Fourier transform, ${}^*$ is the complex conjugate, and $S_n(f)$ is the noise power-spectral density (PSD).  
Eq.~\eqref{eq:SNR} is the loudness in a given detector. When facing a network of detectors, the loudness of the signal is given by the network SNR
\begin{equation}\label{eq:net_SNR}
    \rho_{\mathrm{net}} = \sum_{i = 1}^{N_{\mathrm{det}}} \rho_i^2 \, ,
\end{equation}
where $N_{\mathrm{det}}$ is the number of detectors in the network, and $\rho_i$ is the SNR in detector $i$, calculated according to Eq.~\eqref{eq:SNR}. 

For training the model, it is beneficial to characterize the power of a signal without needing to inject it into noise, which is particularly useful when generating large numbers of signals. Therefore, we characterize signals using the optimal SNR ($\rho_{\mathrm{opt}}$), corresponding to the template matched with itself:
\begin{equation}
    \rho_{\text{opt}} = \left(4 \Re\left(\int_{f_{\text{min}}}^{f_{\text{max}}} \frac{\tilde{h}(f)\tilde{h}^{*}(f)}{S_{n}(f)}  \right)\right)^{1/2} \,.
    \label{eq:optSNR}
\end{equation}
In essence, this means neglecting the noise effects on the SNR. Again, this optimal SNR can be used to compute a network optimal SNR as done in Eq.~\eqref{eq:net_SNR}.

Looking at Eqs.~\eqref{eq:SNR} and~\eqref{eq:optSNR}, we see that they go from $f_{min}$ to $f_{max}$, which would typically correspond to the lowest sensitive frequency of the detector to the maximum frequency reached by the signal, roughly corresponding to the innermost circular stable orbit. However, in this work, we do not look at the full signal but only a part of it. Therefore, we use the \emph{partial inspiral SNR} (PISNR) defined in Refs.~\cite{Baltus:2021nme, Baltus:2022pep}. Essentially, it is defined the same way as the SNRs above (Eqs.~\eqref{eq:SNR} and~\eqref{eq:optSNR}) except the maximum frequency is $f_{cut}$, corresponding to the maximum frequency the inspiral reaches when we are a given time before the merger. In our case, we chose this cut when injecting the signal and training the network to recover the event's characteristics. 

\section{Analysis Setup}\label{sec:analysis_setup}

In this section, we present the various important components of this work. We first present our context network. Then, give a brief explanation of our detection and neural posterior estimation (NPE) algorithms. Finally, we explain which signals are considered in this work.

\subsection{Context Network}
\label{subsec:context_network}

Before detecting or characterizing the events, it is important to convert the data from the detector output into a format that is easily interpretable by the network. This is typically done by a context (or embedding) network. Here, it takes as input the detector strain for each detector in the network---the two LIGO and the Virgo detectors at design sensitivity~\cite{aLIGOdesign, TheVirgo:2014hva}---projected onto an SVD basis~\cite{Cannon:2010qh} and converts them into context tokens which can be passed to other neural components. We pass the strain data in the frequency domain, with a fixed frequency range of 20 to 256 Hz. This approach maintains a fixed input size while allowing us to vary the time before the merger. The latter is done by having a certain maximum frequency reached by the signal, as explained above. Our context network is a residual network~\cite{2015arXiv151203385H} composed of an initial dense layer, 5 residual blocks, and a final dense layer. In this work, we consider signals reaching a maximum value of 40, 70, or 100 Hz. For those, we use 500, 850, and 950 kernels in the SVD decomposition, respectively. By projecting the input to a lower dimensional space, while retaining the signal's information content, the SNR per input dimension increases making it easier for the context network to extract useful patterns from the input. The residual function of each residual block is formed by a multi-layer perception (MLP). The final dense layer compresses the output of the last residual block down to a 256-dimensional vector. This vector will then be passed to the subsequent neural networks used for detection or parameter estimation. 

To ensure this network outputs useful information for the next problem, it is trained at the same time as the NPE network. One could also choose to do it during the detection process. However, we decided to do the former because parameter estimation requires more information than detection, meaning it is likely that if we can infer parameters based on the output, we can also detect signals. 

\subsection{Detection Network}
\label{subsec:detection_network}

The next step is to assess whether a signal is present in the data. Therefore, we pass the context tokens to the detection network. The latter is an MLP trained to classify the input data as either signal or noise. Because it uses the output from our context network, this network does not need to learn the feature extraction and can focus on the detection problem. The MLP has two hidden layers, each with 512 units, and uses the ReLU~\cite{2018arXiv180308375A} activation function. The MLP has a single output node returning a value between 0 and 1, corresponding to the probability of having a signal in the data. Our loss function is binary cross-entropy. To optimize the network, we use the Adam optimizer~\cite{2014arXiv1412.6980K} with the learning rate set to 0.001. With these settings, the network converges within 5 epochs. 

For the training, we use two datasets: one containing the output of our trained context network when a GW signal is present in the data, and one where the analyzed data contains only noise. Both are split evenly, with each comprising 50\% of the 2.000.000 total samples. The data is generated the same way as for the NPE training (see Sec.~\ref{subsec:data_gen}), ensuring consistency in the types of signals the models have been exposed to during training. After training, the MLP's performance is evaluated on a separate test set, also balanced between signal and noise.

Later, to assess our network's capabilities, we look at the true alarm probability (TAP) for a fixed PISNR and a fixed false-alarm probability (FAP). The TAP is the number of signals classified as signals divided by the number of datasets actually containing a GW. The FAP is the number of noise instances classified as a signal by the network divided by the total number of samples containing only noise.  

\subsection{Neural Posterior Estimation}
\label{sec:NPE}

Our last objective is to infer the posterior distribution $p(\boldtheta|d_{obs})$, with $\boldtheta = \{ \mathcal{M}_c, q, \theta, \phi, d_L, \theta_{JN} \}$ corresponding to the chirp mass, mass ratio, sky angles, luminosity distance and inclination, respectively for an observed data $d_{obs}$ containing an inspiraling BNS signal. Conventional methods, such as MCMC or nested sampling, require millions of (sequential) simulations and likelihood evaluations to construct accurate posterior distributions. Under normal circumstances this takes hours~\cite{Veitch:2014wba, Ashton:2018jfp, Romero-Shaw:2020owr}, effectively removing the traditional methods as potential solutions for pre-merger characterization of BNS signals. Modern ML approaches, fueled by advancements in algorithms and the increasing computational power present in GPUs, represent an interesting alternative. Indeed, various ML models to do parameter inference for full GW signals exist~\cite{2019arXiv191202762P, Dax:2021tsq, Dax:2022pxd, Langendorff:2022fzq, Kolmus:2024scm}. Among these models, NPE can estimate the posterior distribution accurately and within seconds~\cite{Dax:2021tsq, Dax:2022pxd, Langendorff:2022fzq}.

NPE learns a mapping from given data $d_{sim}$ to approximate posterior distribution $q_{\boldphi}(\boldtheta|d_{sim})$. The mapping is learned by a neural network with parameters $\boldphi$, which takes as input $d_{sim}$ and outputs the expected posterior distributions. To find the best-fit parameters $\hat{\boldphi}$ for our model, we optimize our framework using Maximum Likelihood Estimation (MLE), corresponding to the following loss:
\begin{equation}
    L = - \sum_{i=1}^n  \log q_{\boldphi}(\theta_{true}^i|d_{sim}^i),
\end{equation}
where $n$ is the number of data points, $\theta_{true}^i$ are the values used to generate the simulated observation $d_{sim}^i$. Using a gradient-based optimizer and MLE we can change $\boldphi$ such that $q_{\boldphi}(\boldtheta|d_{obs})$ resembles $p(\boldtheta|d_{obs})$. Given a sufficiently expressive $q_{\boldphi}(\boldtheta|d_{obs})$, in the limit of infinite data, $q_{\boldphi}(\boldtheta|d_{obs})$ should fit the actual posterior perfectly~\cite{draxler2024universality}.

Any density estimator that can return a likelihood is suitable to model $q_{\boldphi}(\boldtheta|d_{obs})$, the most prominent examples of such estimators are NFs~\cite{rezende2015variational} and mixture density networks~\cite{papamakarios2016fast}. Due to their expressivity and ease of use, we use NFs in our setup. These transform a simple base distribution, like a Gaussian, into a complex target distribution using a sequence of invertible, differentiable functions. Using the change-of-variable rule, one can express the log-likelihood of samples:
\begin{equation}
   \log q(\boldtheta) = \log p(z_0) - \sum_{i=0}^K \log \left| \det J_{f_i}(z_i)  \right| 
\end{equation}
where $p(z_0)$ is our base distribution. We take $f_i$ to be the $i^\mathrm{th}$ invertible, differentiable function in the sequence of transformations, with $z_{i+1} = f_i(z_i)$ and $\boldtheta = z_{K+1}$. $J_{f_i}(z_i)$ is the Jacobian corresponding to the $i^{\mathrm{th}}$ function. By using coupling-layers~\cite{coupling_flow}, the computational load of the log determinant of the Jacobian is manageable while preserving the ability to model complex distributions. 

For our purpose, the NPE takes in the output from the context network and uses it to produce conditioned posteriors. The NF model uses a truncated normal as its base distribution and uses four coupling layers to transform the base distribution into the target distribution. Each flow layer applies a 64-order Bernstein polynomial~\cite{ramasinghe2021robust} as the transformation function, providing high expressivity and robustness. The coefficients for these Bernstein polynomials are generated by a single-layer MLP. To find the optimal neural network weights we use the Adam optimizer~\cite{2014arXiv1412.6980K} with its learning rate set to 0.0001.  It takes between four and five days to fully train a network, using an Nvidia A100-PCIE-40GB GPU. Multiple runs were conducted to find the optimal set of hyperparameters.

\subsection{Data Generation}
\label{subsec:data_gen}

\begin{figure}
    \centering
    \includegraphics[width = 0.5\textwidth]{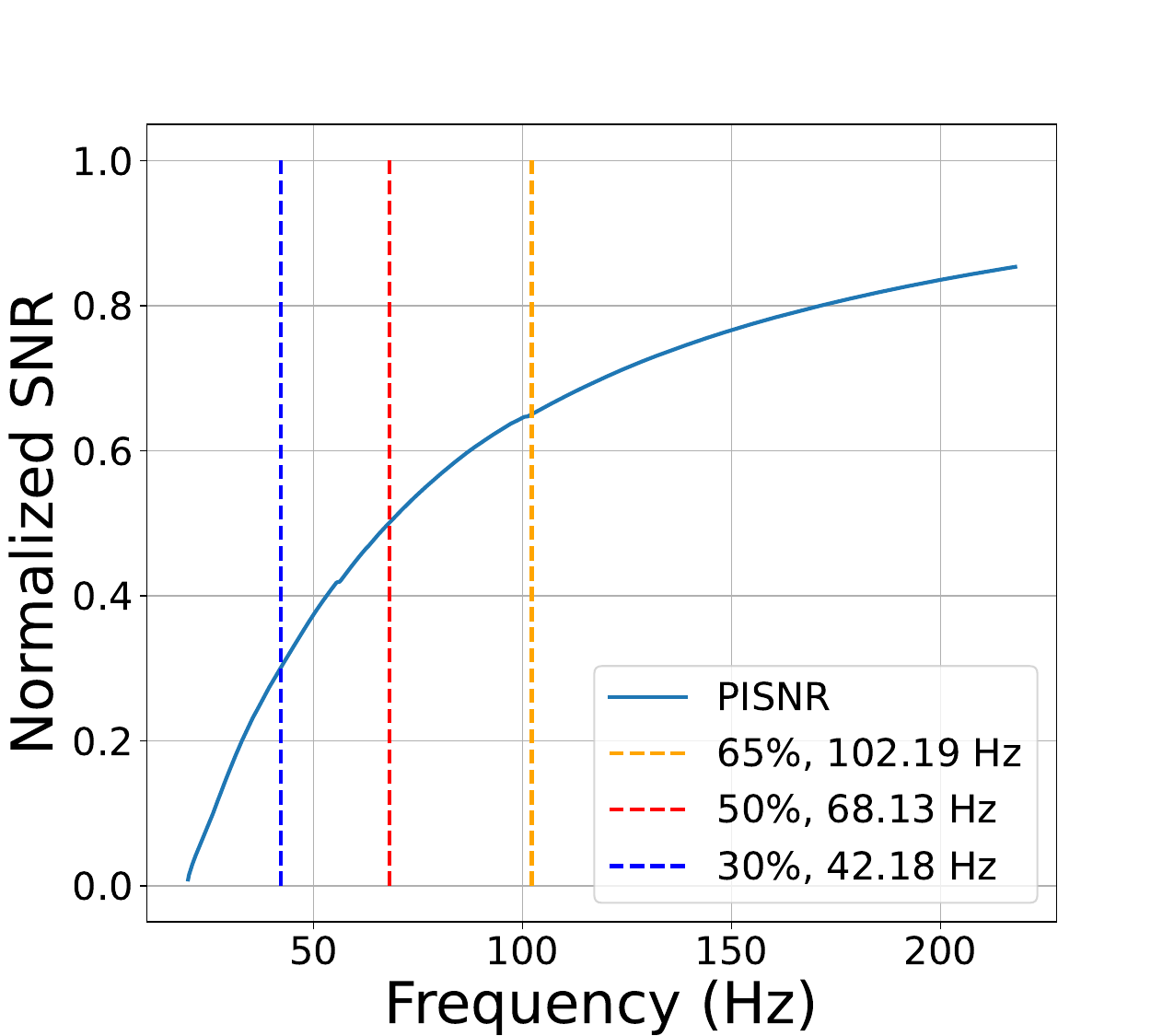}
    \caption{Fraction of the total SNR contained in the PISNR as a function of the maximum frequency of the signal. We train the networks at 100 Hz, 70 Hz, 50 Hz and 40 Hz cut frequencies.}
    \label{fig:PISNR}
\end{figure}

As mentioned previously, we consider a detector network made of the two LIGO detectors, and the Virgo interferometer at design sensitivity~\cite{aLIGOdesign, TheVirgo:2014hva}. We generate data with a fixed frequency range (20 to 256 Hz) containing a GW with a variable maximum frequency (40, 70, or 100 Hz), typically corresponding to a given time before the merger but also a given fraction of the SNR of the entire signal, as represented in Fig.~\ref{fig:PISNR}. As can be seen from Eq.~\eqref{eq:freq_evolution}, the evolution depends on the mass of the signal. Therefore, this design choice means that we do not fix the time before the merger at which we characterize the signal. However, based on the inferred characteristics, it is possible to assess the duration left quickly using Eq.~\eqref{eq:time_freq_evolution}. 

We inject our signals using the \textsc{IMRPhenomD\_NRTidalv2} waveform~\cite{Dietrich:2019kaq} as coded in the \textsc{Ripple} package~\cite{Edwards:2023sak}, relying on \textsc{JAX}~\cite{jax2018github}. A summary of the distributions used to generate the events is given in Table~\ref{tab:priors}. We take a uniform distribution in chirp mass with masses ranging from $1.13\,M_{\odot}$ to $2.61\,M_{\odot}$. The distribution for the mass ratio ($q=m_2/m_1$) is also uniform, ranging from 0.25 to 1. For the distance, we do not select it from an astrophysical distribution but rather select the PISNR of the signal from a beta distribution ranging from 5 to 50, with a temperature of 10 and a peak of 15. This ensures the signals are loud enough to be detected by the main searches, whether done classically or with ML studies~\cite{Sachdev:2020lfd, Baltus:2021nme, Baltus:2022pep}. Note, however, that we do not put any constraint on the individual (PI)SNR in each detector, contrary to what has been done in previous approaches~\cite{Sachdev:2020lfd, Chatterjee:2022dik}. The other parameters are sampled from their usual prior distributions. With these parameters and the chosen frequency range, our system can provide sky maps up to one minute before the merger for current detectors. This can naturally be improved as interferometers grow more sensitive in the low-frequency regime. 

\begin{table}[t]
\begin{tabular}{|c|c|}
    \hline
    Parameter & Distribution \\
    \hline 
    \hline
    Chirp mass $\mathcal{M}_{c}$ & Uniform ($1.13,2.61$)\\
    Mass ratio $q$ & Uniform ($0.25,1.0$)\\
    Spin amplitudes & Uniform ($0.0,0.5$)\\
    Inclination angle $\theta_{JN}$ & Uniform ($0,\pi$)\\
    $\text{PISNR}$ & Scaled Beta (5,50) \\
    \hline
\end{tabular}
\caption{Table showing the main distributions used to generate the signals. The beta distribution has a temperature of 10 and a peak of 15. The beta distribution is scaled to have a minimum of 5 and a maximum of 50. The other parameters are samples from their usual prior distributions.}
\label{tab:priors}
\end{table}

During the training of the context and NPE networks, we generate batches of 1000 signals to train the network. Each epoch contains 9 training batches and 1 validation batch. We train the network for about 4000 epochs. To improve the network's sensitivity, we use \textit{curriculum learning} on the PISNR. The general idea behind it is to start the training with easier signals, corresponding to louder signals, and ramp up the difficulty over time, going to lower PISNR signals. Therefore, we change the PISNR distribution from a beta distribution with a temperature of 35 and a peak of 40 to the desired temperature of 10 and a peak of 15. The maximum and minimum of the beta distribution stay at 5 and 50 throughout the process.

\section{Detection Capabilities}
\label{sec:detection_results}

\begin{figure}
    \centering
    \includegraphics[keepaspectratio, width = 0.5\textwidth]{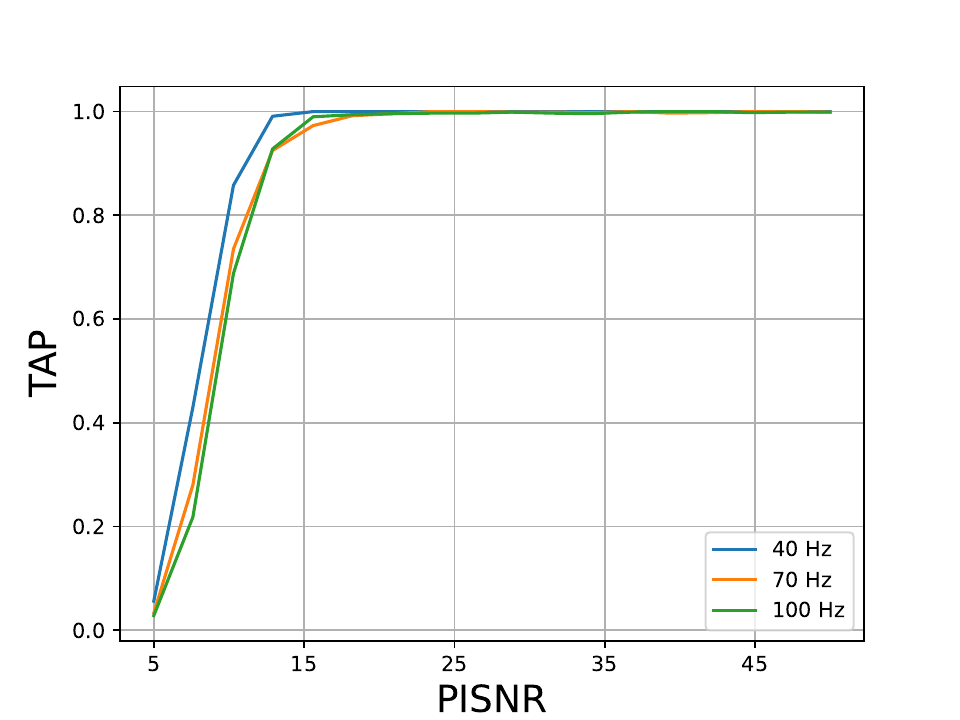}
    \caption{True alarm probability as a function of the PISNR for the 40, 70, and 100 Hz cut frequency for a fixed false-alarm probability of 1\%. We see that the PISNR curve is about the same for all the maximum frequencies. It is slightly better for the 40 Hz case, which we expect is due to shorter signals leading to fewer SVD components and increasing the information content per SVD kernel. Still, these results translate to a better detection for higher frequencies in practice since the SNR accumulates as more of the signal enters the detector.}
    \label{fig:TAP}
\end{figure}

In this section, we discuss the results regarding our detection algorithm and the experiments done to assess our detection capabilities. To do so, we make a data set of signals distributed with various PISNRs in the range of 5 to 50 and see how many are detected and how many are missed as a function of the PISNR. We represent the result in Fig.~\ref{fig:TAP} with a fixed FAP of 1\%. The TAP dependency is close for all the frequency cuts as a function of the PISNR. This is somewhat similar to what was observed in Ref.~\cite{Baltus:2022pep} for the highest frequencies considered by the authors. We also note that we see a slight improvement for the lowest frequency. This is a consequence of the use of the SVD decomposition and a fixed PISNR. Indeed, for a lower maximum frequency, the amplitude of the signal compared to the noise is increased to keep the same loudness. Therefore, the information per frequency bin is increased. This also means that there is more information per SVD kernel. We expect this to be the cause of the difference. Still, this does not mean more events would be detected at lower frequencies because, for a realistic observation, there is also an SNR accumulation effect, meaning the same signal at 100 Hz is always louder than it is at 40 Hz. This can also be seen in Fig.~\ref{fig:PISNR}. So, it is the most likely one detects signals with higher frequency cut-offs.
Assessing the difference in efficiency between our detection algorithm and the ones already present in the literature~\cite{Sachdev:2020lfd, Baltus:2021nme, Baltus:2022pep, Yu:2021vvm} is not easy since all of these works use different setups, i.e. they have different PSDs but also different cut frequencies. One main observation is that our results may look quite optimistic compared to others but we also use higher cut frequencies compared to others. More efforts to go to low frequencies would be needed to do a fair comparison and to have more latency. However, here, we decided not to go too low in frequency to ensure we could have reasonable sky maps even for the lower cut frequency, hence keeping enough information content in the signals. Exploring lower frequencies is left for future work. 

\begin{figure}
    \centering
    \includegraphics[keepaspectratio, width=0.5\textwidth]{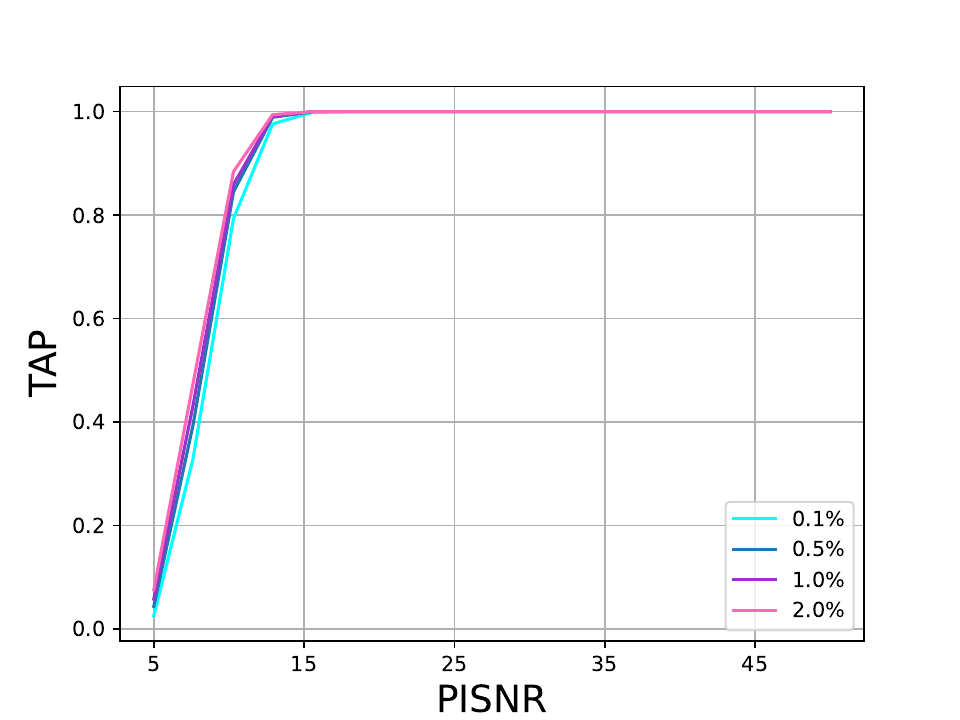}
    \caption{True alarm probability as a function of the PISNR for different false-alarm probabilities for the network trained with a 40 Hz maximum frequency. While a decrease in performance is seen, it is not dramatic either, meaning we could reduce the number of spurious triggers without losing too many detection capabilities.}
    \label{fig:tap_change_with_FAP}
\end{figure}

It has also been shown in past work that a high FAP can lead to many triggers if run on a realistic search~\cite{Baltus:2021nme, Baltus:2022pep}. Therefore, it is also interesting to investigate the evolution of the TAP as we change the FAP. For a given network---here taken as the one with a 40 Hz maximum frequency though the results are equivalent in all cases---, we look at the evolution of the TAP curve for different FAPs. The results are shown in Fig.~\ref{fig:tap_change_with_FAP}. While there is a decrease in the detection efficiency it is not extreme either. In particular, the change comes mostly in the region where the TAP evolves, leading to a slight reduction for SNRs below 20. Above that, events are always detected for all the FAP tested. Still, this corresponds to signals with extremely high SNRs.\footnote{This is true for the 40 Hz case. For higher cut frequencies, SNRs can be more moderate.} So, improving detection efficiency while keeping FAP low in the low SNR region is important. Several strategies have been proposed in the literature. In Ref.~\cite{Baltus:2022pep}, they suggest requiring several triggers in a row when more signal enters the detector. This reduces the latency but removes many false alarms. We could do the same by requiring detection in several networks trained for different maximum frequencies as time goes on. On the other hand, we could also adapt our full framework to work with a variable maximum frequency for the signal and do the same trick. Alternatively, in Ref~\cite{Yu:2021vvm}, they use a denoiser to boost the detection probability. Such an approach could also be undertaken here. We leave this for future work. 

\section{Inference Results}
\label{sec:npe_results}

In this section, we look in detail at the results found for the inference problem. In principle, we should only look at the results for events that have been detected. However, here, to investigate our NPE's performance, we consider all events that fall within the priors used for training.

\subsection{Accuracy of our framework}

\begin{figure*}[h]
    \centering
    \includegraphics[keepaspectratio, width = 0.8\textwidth]{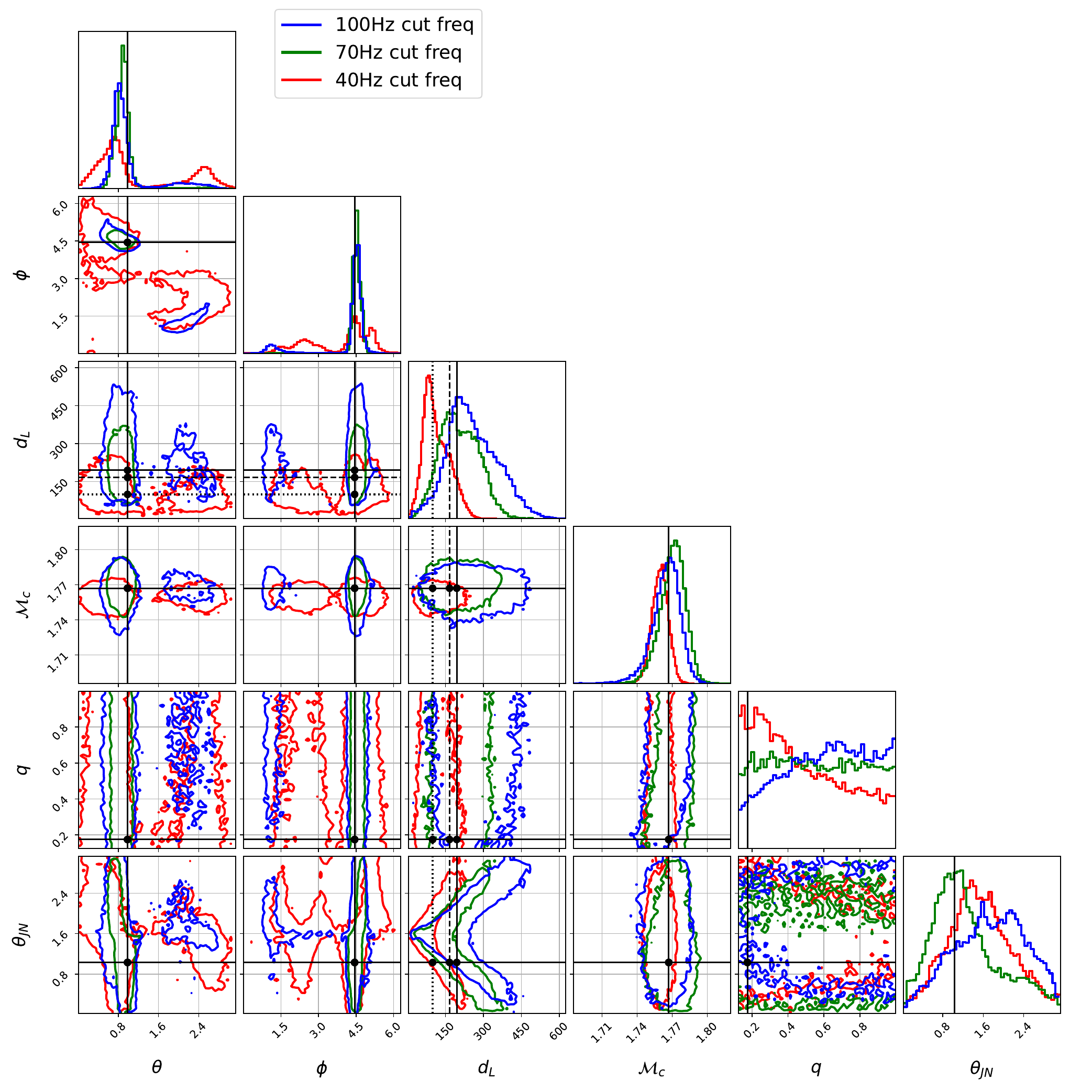}
    \caption{Corner plot representing the parameters recovered by our approach for an inspiraling BNS for different maximum frequencies but scaled to keep a constant PISNR of 13.5. For all the maximum frequencies, our setup properly recovers the injected values, even if the posteriors become broader for lower maximum frequencies, as expected based on the decrease in the number of signal cycles present in the data and from which information can be retrieved. In the luminosity distance estimation, we plot the true values for the networks separately since they have different luminosity distances to keep a constant PISNR. The dotted line is the true value for the 40 Hz network, the dashed line is the true value for the 70 Hz network and the solid line is the true value for the 100 Hz network. All the other parameters are the same for all network.}
    \label{fig:cornerplot}
\end{figure*}

First, we look at the posteriors resulting from the analysis of a given signal. An example of this is given in Fig.~\ref{fig:cornerplot}, representing the posteriors obtained for different maximum frequencies and fixed PISNR. The latter is done to avoid the broadening of the posterior due to the PISNR decreasing when we decrease the maximum frequency reached by the signal. Nevertheless, a reduction in accuracy is observed for the lower frequencies. This is expected since the network observes fewer cycles in band and has less information to fit the signal. Still, in all cases, our approach properly recovers the injected parameters, whether it is the sky angles or other parameters relevant to the EM follow-up. 

\begin{figure*}
    \centering
    \includegraphics[keepaspectratio, width = 0.49\textwidth]{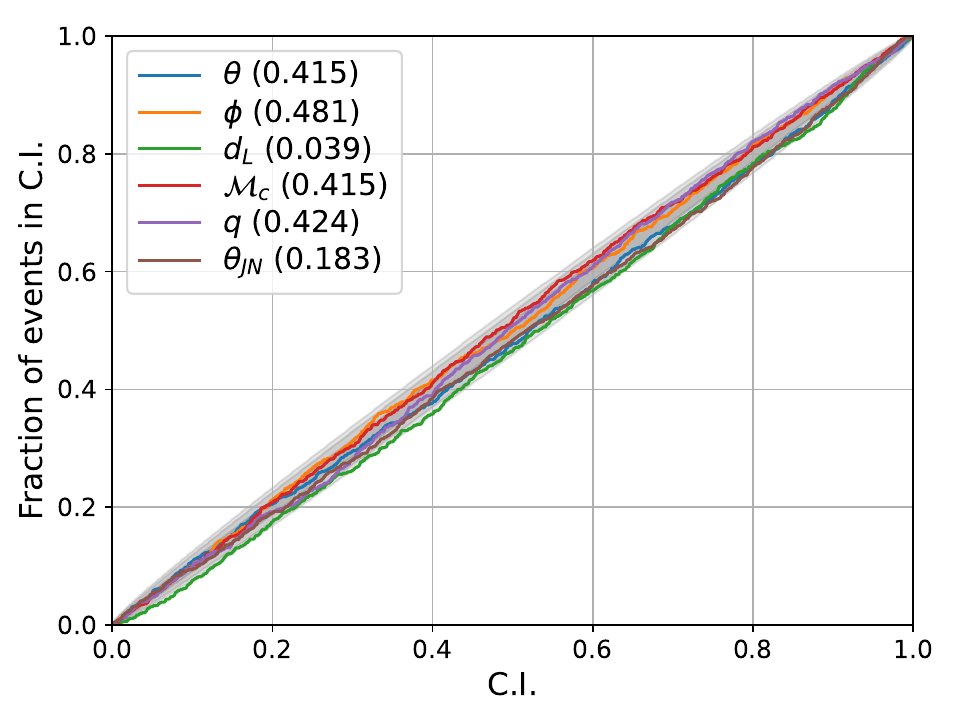}\includegraphics[keepaspectratio,width = 0.49\textwidth]{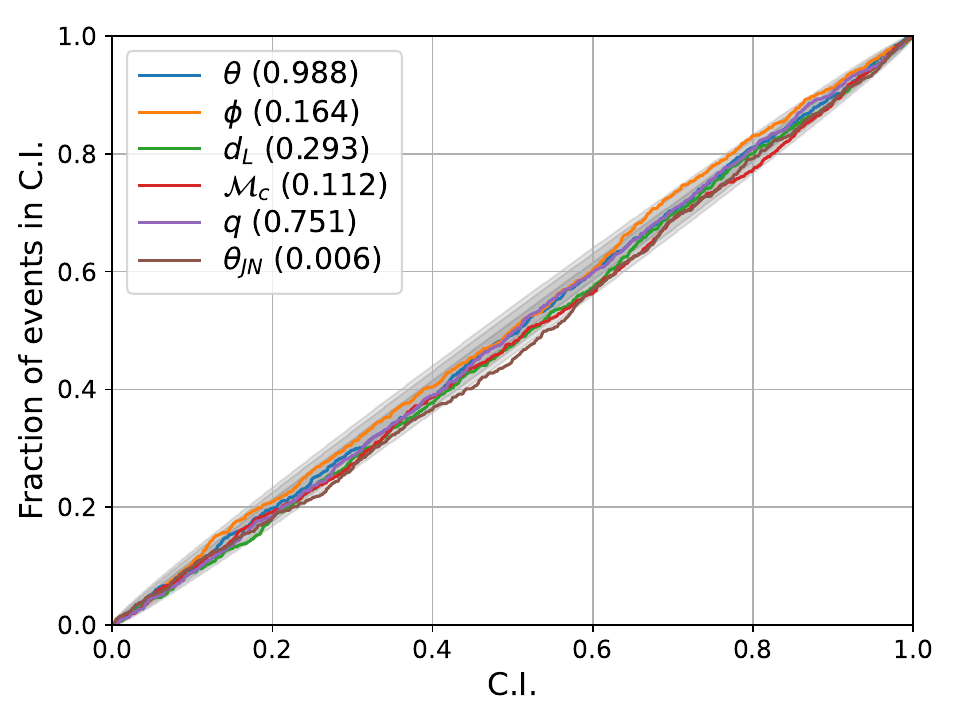}
    \caption{PP plots for our inference framework with a maximum frequency of 40 Hz (left) and 100 Hz (right). The values in brackets represent the p-values for the different parameters. In both cases, we follow the diagonal quite closely.}
    \label{fig:pp-plots}
\end{figure*}

To verify the stability of the results, we generate probability-probability plots (PP-plots), as shown in Fig. \ref{fig:pp-plots}. Such plots translate the fraction of injections for which the true value falls within a given confidence interval, and it should be a diagonal if the network is not biased. We note, however, that this test does not guarantee a perfectly-functioning network but it is a necessary condition to be unbiased. One sees the results are unbiased for the two chosen cut frequencies and all the parameters estimated by our network. This is also the case for the other cut frequencies on which our networks are trained. This means we have the expected statistical behavior on a large number of events.

\subsection{More detailed investigation of the inference results}

The probability of observing the EM counterpart of a BNS system also depends on the luminosity distance and inclination (if the event is too far, its apparent magnitude on Earth may be such that we cannot see the EM counterpart and the jet emitted is expected to be beamed perpendicular to the orbital plane, and higher chances of detection happen when this is more pointed towards the Earth), but the most critical piece of information to start a follow up is the sky localization, which determines where the EM telescopes need look and where the EM counterpart should be located. If the map is too broad, one cannot possibly cover the full area rapidly and transient signals may be missed. Moreover, it increases the chance of having an EM emission that is unrelated to the GW event we just observed. So, it is instructive to analyze how the posterior distribution for the localization changes depending on the maximum frequency (and thus the time before the merger) and the signal PISNR.

\begin{figure*}
    \centering
    \includegraphics[keepaspectratio, width = 0.49\textwidth]{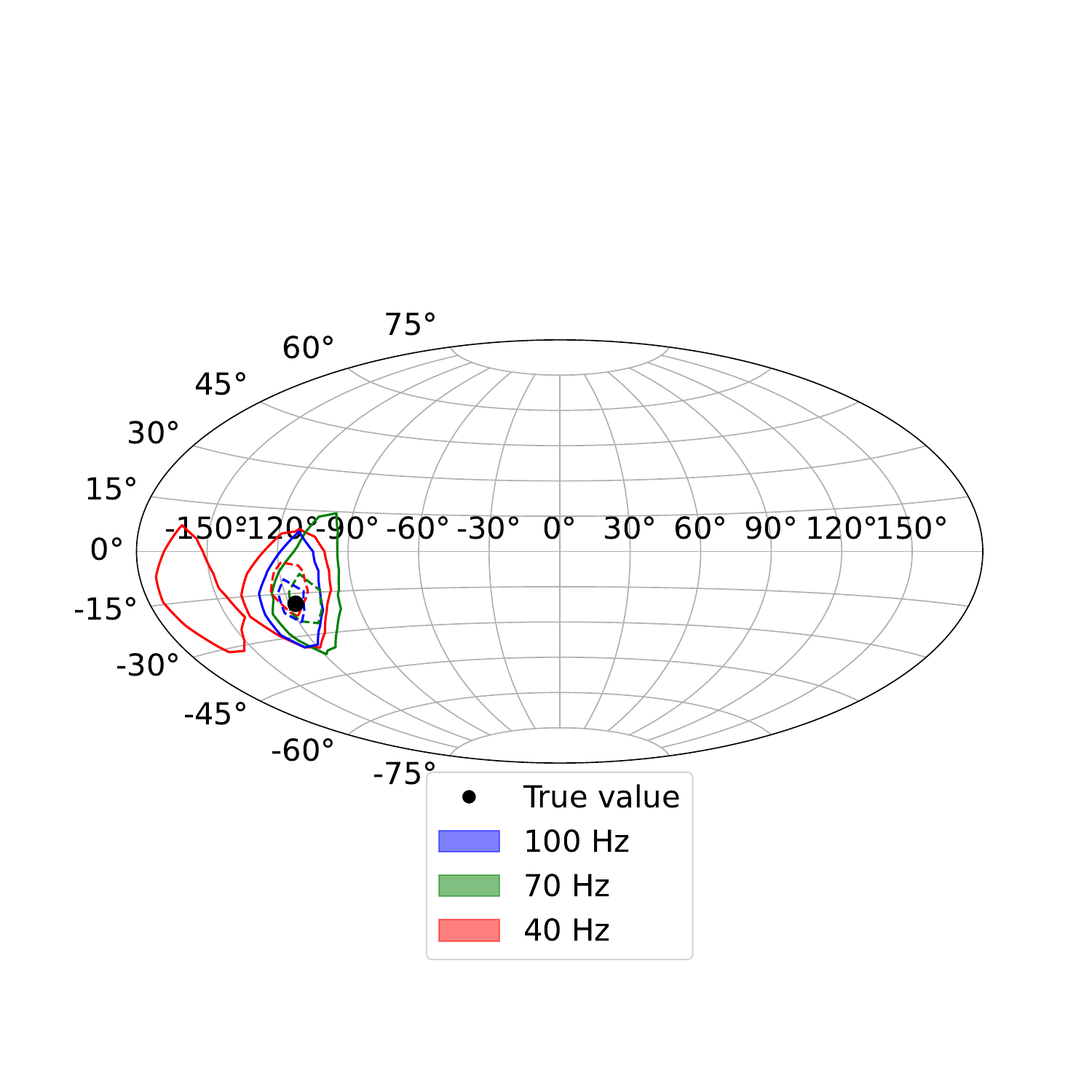}
    \includegraphics[keepaspectratio,width = 0.49\textwidth]{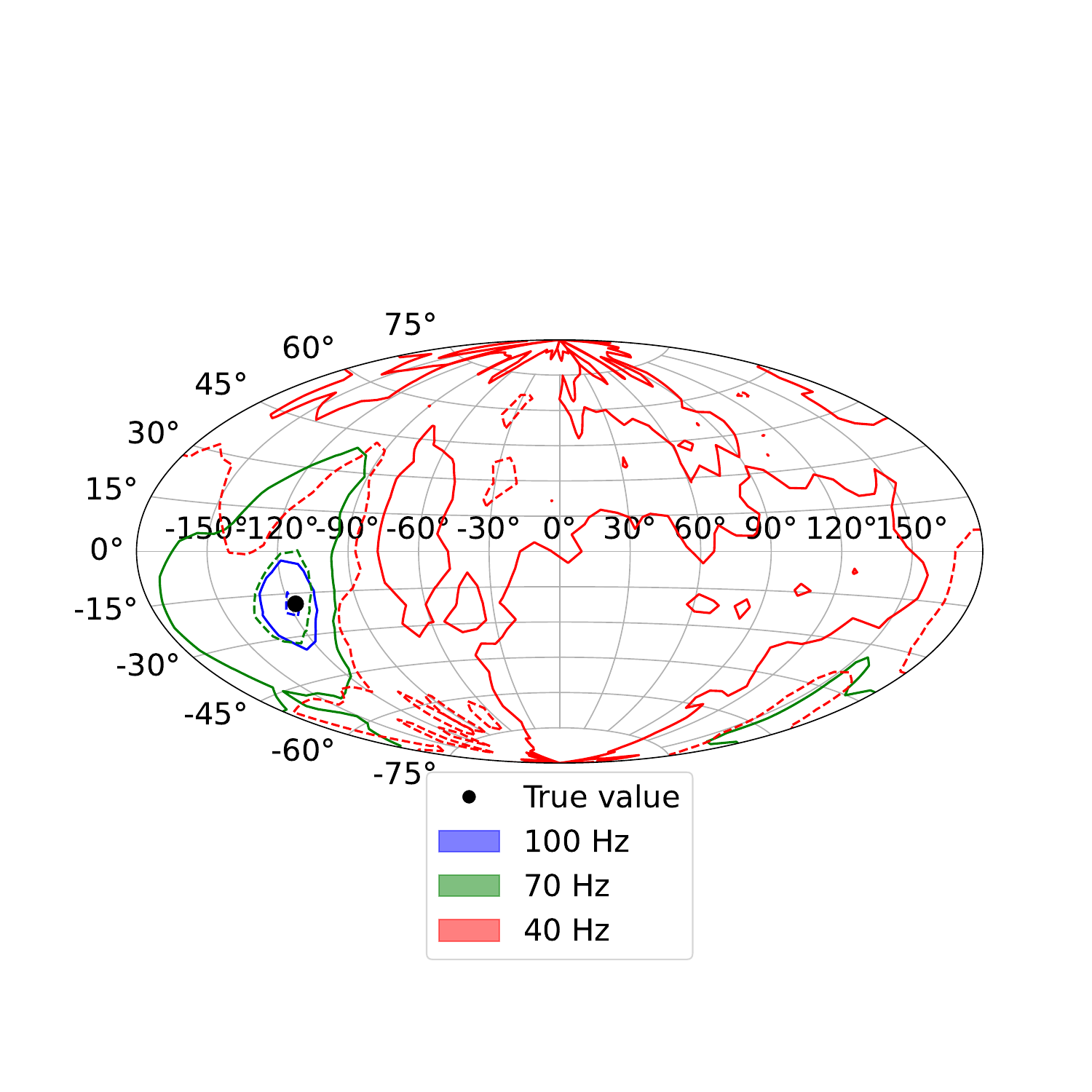}
    \caption{Example sky maps obtained with our framework. The left figure is at a constant PISNR of 17.1. It shows that the networks with a lower maximum frequency perform worse than those with a higher maximum frequency due to the network observing fewer signal cycles in the sensitivity band. The right figure shows an evolving sky map in a more realistic scenario where all the parameters are kept the same from one situation to the other. In this case, the PISNR is 8.9, 14.6, and 17.1 for a maximum frequency of 40, 70, and 100 Hz, respectively. Here, the difference from one maximum frequency to the other is bigger as the posterior degrades due to fainter signals occurring for lower maximum frequencies.}
    \label{fig:evolving_skymaps}
\end{figure*}

First, as an example, Fig.~\ref{fig:evolving_skymaps} shows two sky maps for different maximum frequencies reached by the signal. In the left panel, all parameters are the same for the three cases, except the luminosity distance, which is scaled to keep a constant PISNR of 17 for all cut frequencies. In this case, one sees that the network provides reasonable sky maps every time. As expected, the sky location improves when a larger maximum frequency is seen since more signal cycles are in band, enabling one to better capture the dephasing, time delay, and PISNR difference between the detectors to reconstruct the origin of the signal. The same improvement is observed for the other parameters. For the case where the PISNR is not kept fixed but instead scaled with a constant luminosity distance ---corresponding to a more realistic case---, the sky map evolves logically, with the largest estimated sky area for the lowest maximum frequency, also corresponding to the lower PISNR. The effect of lower frequency was discussed before and is here made worse by the lowering of the PISNR, where a quieter signal generally leads to worse results (all other parameters being fixed). As a consequence, while our network behaves as expected, it also implies that the obtained sky maps may not be good enough to actually locate the host before it merges in a part of the realistic scenarios. This was also seen in other studies done with different approaches~\cite{Sachdev:2020lfd, Chatterjee:2022dik}. For example, for the PISNR of 17 at 40 Hz, the total SNR of the signal would be about 50. Such a high SNR has not been observed yet but could be expected when the detectors get upgraded, showing the importance of following up for early alert in the future since, at least for the loudest examples, the sky map is good enough to warrant an EM follow-up. 

\begin{figure*}
    \centering
    \includegraphics[keepaspectratio, width = 0.49\textwidth]{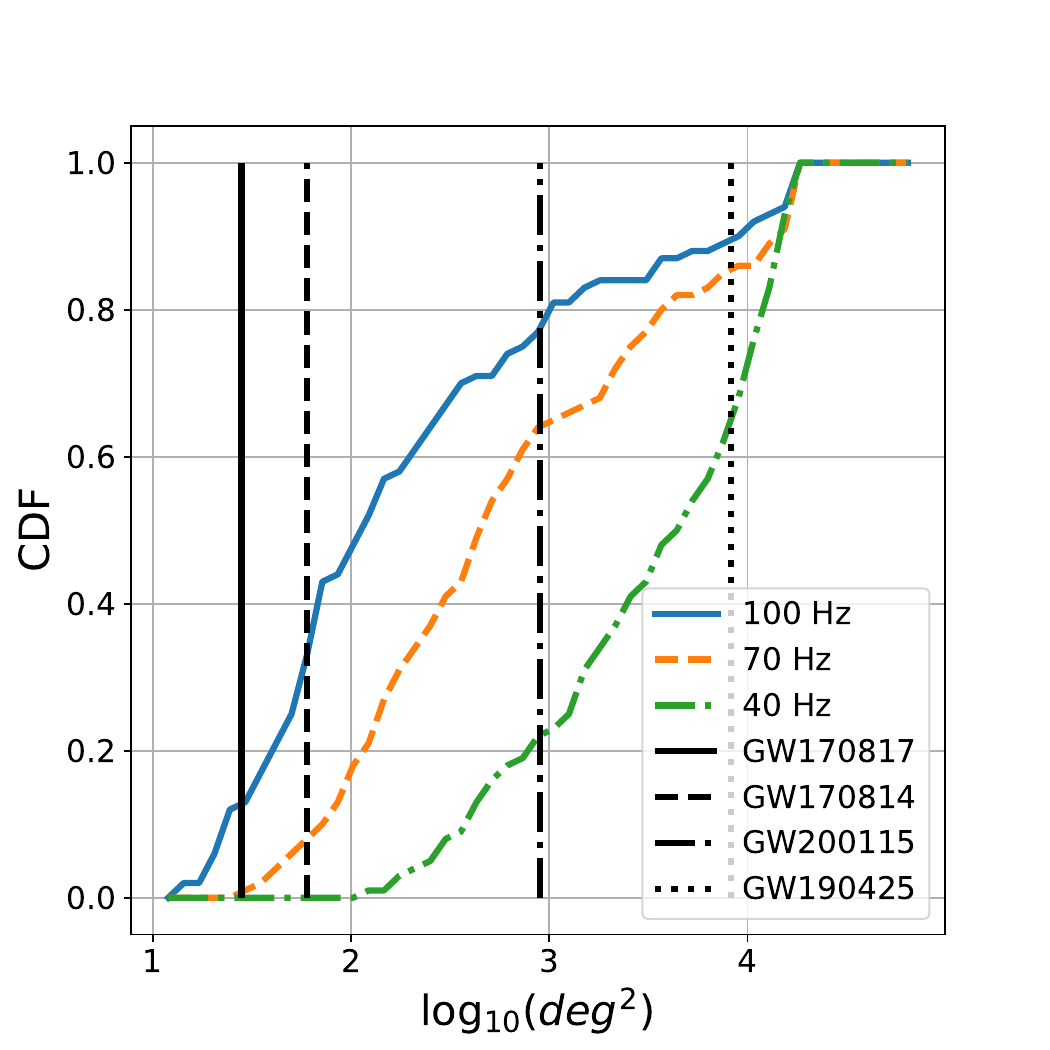}
    \includegraphics[keepaspectratio,width = 0.49\textwidth]{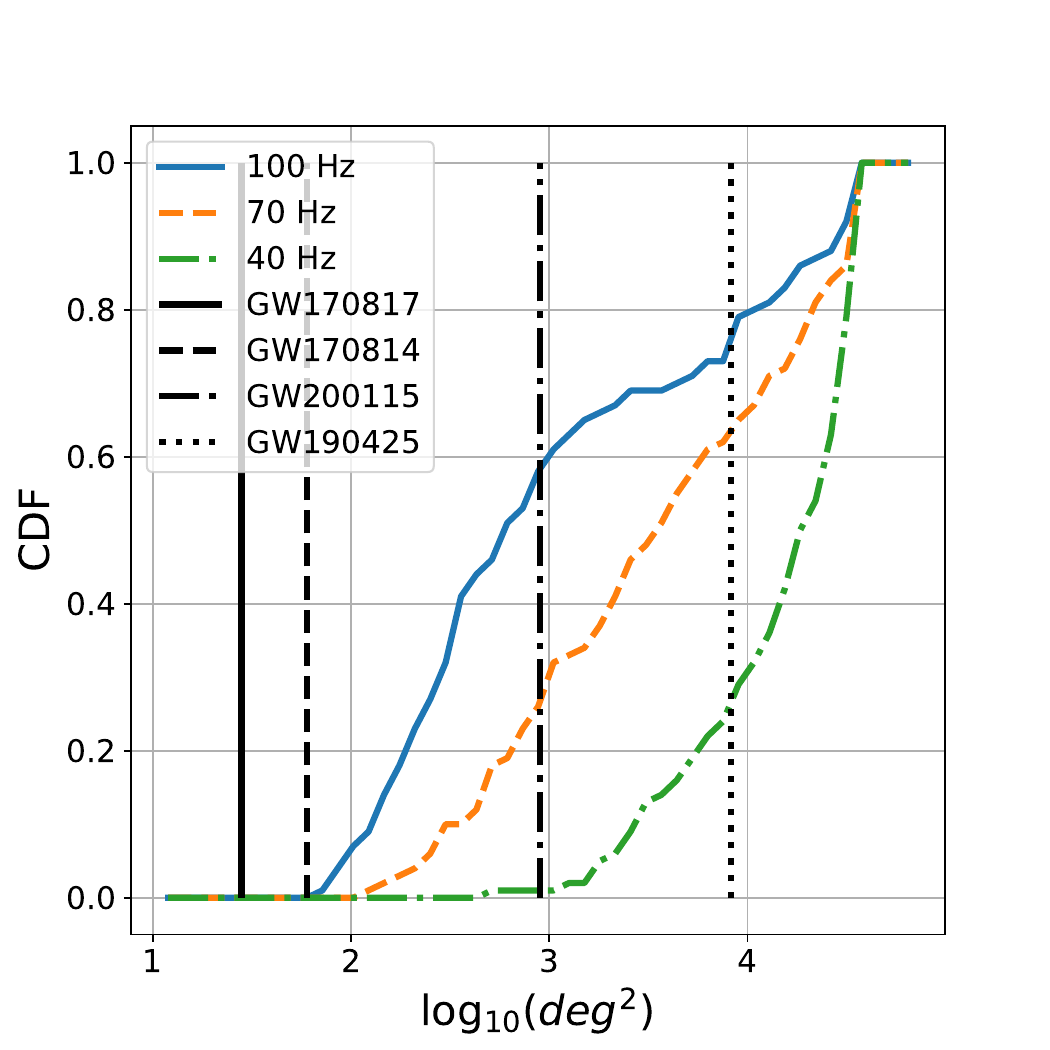}
    \caption{Cumulative density functions for the log of the estimated sky area of the different networks. The figure on the left shows the sky area of the 50\% confidence interval, and right one shows the 90\% confidence interval. For reference, we also represent the 90\% confidence area for some reference GW events. The general trend is that the networks with a higher cut frequency perform better than the networks with a lower cut frequency. Some cases are already constrained enough to offer interesting perspectives for EM follow-up.}
    \label{fig:sky_hist}
\end{figure*}

Besides these examples, we can also check the localization capacities of our approach on a larger scale. Therefore, we take 1000 events and look at their corresponding sky area at 50\% and 90\% confidence intervals when they reach a given maximum frequency. The results are shown in Fig.~\ref{fig:sky_hist}, where we show the cumulative density function of the area distribution. The results obtained match those found with other approaches~\cite{Chatterjee:2022dik}, though we stress our approach is different and also encompasses other parameters since it does not rely on matched filtering information at all. Therefore, the expectations for EM follow-up are closely related to those already proposed in the literature: for the best signals (loudest and well-detected in all detectors), there are interesting perspectives on the possible localization of BNS signals before their merger. However, to make this more frequent, more developments on the detectors are needed, with a better overall sensitivity and lower-frequency reach potentially boosting the prospects significantly. Nevertheless, our approach here shows that NF-based frameworks are a promising avenue to parse all the information needed for deciding whether a signal warrants EM-follow-up in the coming years, which is in line with the various successes of the method to do parameter estimation tasks~\cite{Dax:2021tsq, Dax:2022pxd, Kolmus:2024scm}. 

In the end, this shows our network produces reliable and reasonable posterior for all signals within the training priors. However, even if we can detect very faint signals, the information content may not be sufficient to actually have a sky map permitting a good pre-merger EM follow-up. Still, knowing the signal is coming and preparing for EM observation as soon as a sky map is available would be interesting, especially since no prompt emission at the merger or close to it has ever been observed.

\section{Combined Detection and Characterization of Inspiraling BNS Systems}
\label{sec:joint_results}

\subsection{Mock Observation Scenario}

To illustrate the detection and characterization capacities of our framework, we look at simulated signals, find which one can be detected, and their inferred posterior distributions. Therefore, we follow the same approach as in Refs.~\cite{Baltus:2021nme, Baltus:2022pep}, simulating the BNSs one would detect in 5 years of observation for the LIGO-Virgo network at design sensitivity. The code used for the generation is the same as in Refs.~\cite{Samajdar:2021egv, Baltus:2022pep}, with the adaptation that the local BNS merger rate follows the latest LVK results~\cite{KAGRA:2021duu}. We also changed the PSDs to follow the ones used throughout this work. Here, we only consider the detected signals, assumed to be those with a total SNR larger than 8. However, we note that, since our networks have been trained on faint signals, they could have the potential to also look for sub-threshold signals, though we would need to adapt the network and ensure a low false-alarm probability. With our setup, we find 90 BNS mergers that can be detected over the course of five years when the complete signal is in band.

For each of these signals, we make a data frame to pass the data in the context network when they reach 40, 70, and 100 Hz. The output is then used to see if the events are detectable for different maximum frequencies. We find there are 17 signals we can already detect at 40 Hz, 24 signals that need to reach 70 Hz before being detectable, and 32 signals that require 100 Hz. This also means that 58 signals cannot be detected in advance with our method. However, those signals have low total SNR (close to 8), meaning they enter the lower SNR end of the training distribution\footnote{Some of the signals even have an SNR below the lowest bound used when training.}, where the TAP, which can be seen as the detection probability, is low (see Fig.~\ref{fig:TAP}). These numbers are also in line with others found in the literature. For example, in a similar exercise, \emph{Baltus et. al}~\cite{Baltus:2022pep} detect 15 signals out of 100 in advance with maximum frequencies reached by the signal ranging from $\sim 25$ Hz to $\sim 65$ Hz, noting that in their experiment, there is no strict limit on the frequencies required. From this, it seems like our detection network is doing slightly better, although it detects part of the signals later, which is a drawback from the low-latency perspective. Still, this gives interesting perspectives for our approach, which could be improved to also work for lower frequencies in the future.

\begin{figure}
    \centering
    \includegraphics[keepaspectratio, width = 0.5\textwidth]{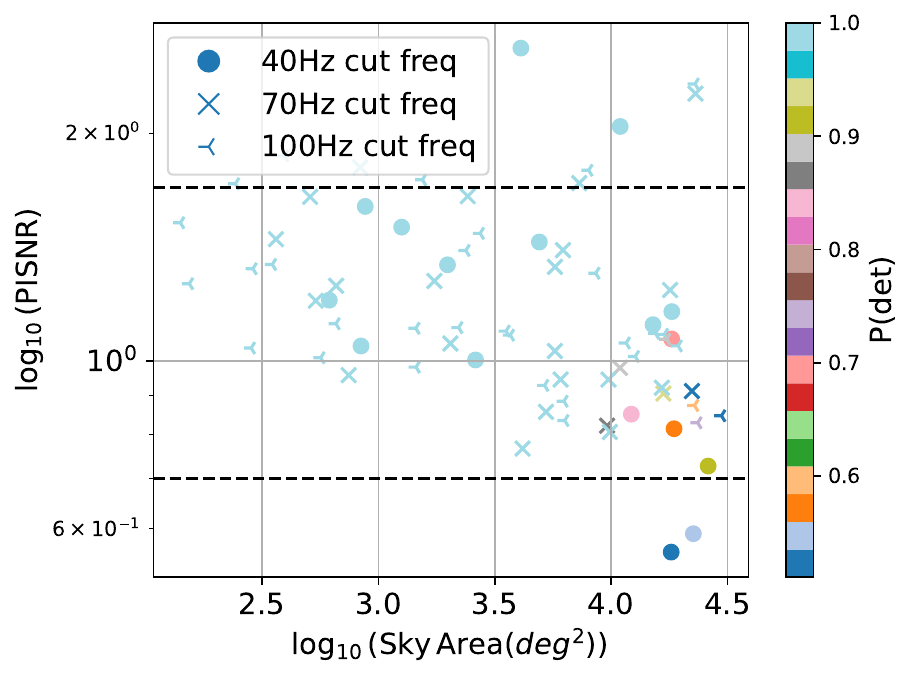}
    \caption{Representation of the PISNR versus the 90\% confidence sky area for our detected events. The color scheme shows the detection probability for the various events. We only consider events that would be detected. The dashed line represents the PISNR range on which our network has been trained. Therefore, points lying outside are outliers and the sky estimates cannot really be trusted. Within the trained region, we see the expected behavior, with a better sky location and detection probability for the higher PISNR events and higher maximum frequencies.}
    \label{fig:realistic_observations}
\end{figure}

For the detected events, we pass the context information to the NPE network. We represent the result of the sky areas found for the different events as a function of their PISNR and detection probability in Fig.~\ref{fig:realistic_observations}. First, we note that some of the simulated and subsequently detected events are outside of our training regime. Therefore, related results cannot really be trusted. This shows our context network is relatively robust compared to the PISNR it is trained on but also that our PISNR range should be widened before applying our framework for real searches and characterizations. This would require longer training. Nevertheless, for the events that are detected and inside the training region, we see a clear hierarchy; louder signals have a better detection probability and sky map for a given maximum frequency, and a higher maximum frequency leads to better detections and sky maps for a given PISNR. This is also in line with results shown in other sections in this work, in particular, Fig~\ref{fig:sky_hist}. This shows our sky localization framework behaves as expected on those signals. Unfortunately, the accuracy of these sky maps is not always great. For the 70 Hz cut, we have only six cases with an area for the 90\% confidence region below $1000 \, \mathrm{deg}^2$, with the best sky map having an area of $363 \, \mathrm{deg}^2$. For the 100 Hz case, there are 9 cases with an area below $1000 \, \mathrm{deg}^2$ for the 90\% confidence region, with 5 having an are below $500 \, \mathrm{deg}^2$. From these, the best-localized event has an area of $141 \, \mathrm{deg}^2$. In this case, one could envisage doing EM follow-up with a wide field of view telescopes, though there is no guarantee to find the signal early enough. The main milestone required to really do pre-merger alerts in this case is to have detection and good accuracy at earlier times, requiring us to be sensitive to lower frequencies. Perspectives would naturally improve for lower minimum frequencies, corresponding to more sensitive detectors.

While not being a full pipeline and requiring some more development to be more robust and accurate, the results shown in this section illustrate that our system can naturally be used as a detection and characterization algorithm for inspiraling BNS signals. 

\subsection{Analysis of Realistic Events}

As a final illustration of our framework's capacities, we look at two realistic cases in more detail, corresponding to past observations. On the one hand, we look at a GW170817-like event~\cite{LIGOScientific:2017vwq}, where we take the median values reported in the paper but rescale the luminosity distance to have the same network SNR as the one reported fro detection in the paper. On the other hand, we analyze a GW190425-like event~\cite{LIGOScientific:2020aai}, where we also inject the median values reported in the paper but do not rescale the luminosity distance. Therefore, this event has a larger SNR than the one reported before. Both signals are injected in noise generated from the same PSDs as the ones used above. 

For the detection, the GW170817-like signal can be detected in the 70 Hz maximum frequency network, while the GW190425-like one can already be detected from 40 Hz onwards. This shows that for realistic signals, our detection scheme would be able to find signals beforehand.

For parameter estimation, we look at the inference results for all the cut-off frequencies, regardless of detection as to also compare the inference results with the detection triggers. We give a complete sky map for the GW170817-like signal in Fig.~\ref{fig:GW170817_like}, while the corresponding figure for the GW190425-like signal is given in Appendix~\ref{app:results_GW190425_like}. Regarding the sky map, based on the 90\% confidence interval, EM follow-up would be possible only once the GW170817-like signal reaches 100 Hz, while it would be possible from 70 Hz onwards for the GW190425-like signal. The main reason for this difference compared to what was observed in reality is the scaling we used, meaning that the April 2019 signal has a higher SNR. Moreover, here, it is detected by three detectors, while it was seen by only two in reality.\footnote{LIGO-Livingston and Virgo, though the trigger was below the detection threshold in the latter~\cite{LIGOScientific:2020aai}.} Regarding the other parameters, we observe the same trend as the one reported before: the posteriors improve as more of the signal enters in the detection band, which is expected. For the chirp mass, it is worth noting that the posterior for the GW19425-like signal has some residuals left for the 40 Hz case. Since detection is possible from the features, this indicates our feature network is capable of representing the main features of the signal well. However, the NPE is not confident about the exact posteriors. This could be due to the low information content of the detected signal, and even more exposure to such signals could further improve the inference's robustness~\cite{Kolmus:2024scm}. Nevertheless, providing these sets of samples to the EM observers would enable them to choose the best follow-up strategy since they would know where and what to look for. However, because of the pre-alert nature of the signal, they would also need to deal with the large uncertainties existing on the parameters until the signal can be fully analyzed. 

\begin{figure*}
    \centering
    \includegraphics[keepaspectratio, width=0.9\textwidth]{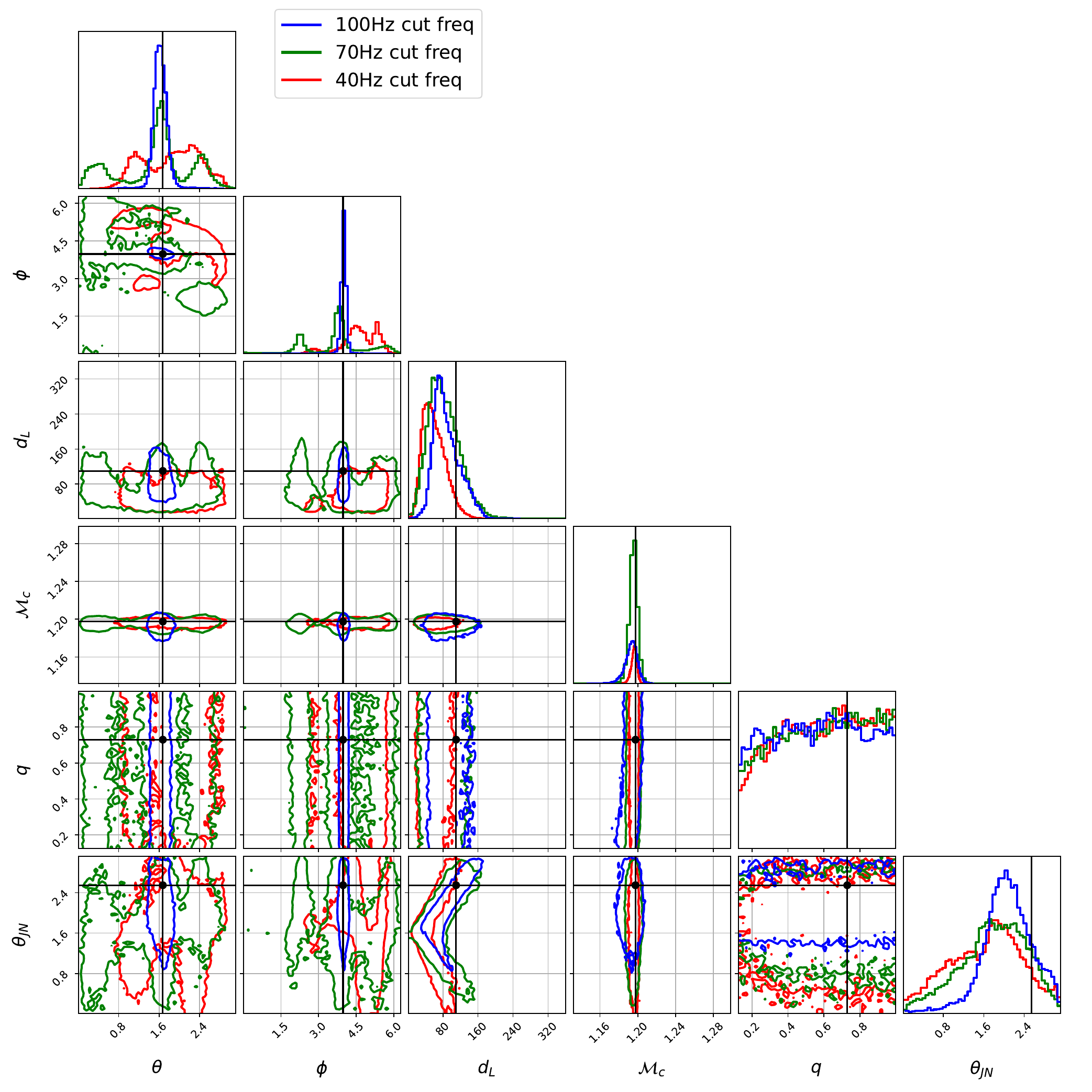}
    \caption{Corner plot for the GW170817-like injection. We find reasonable posteriors for all the networks. However, only when the 100 Hz maximum frequency is reached do we get nice posteriors for the sky localization. For the sky location, EM follow-up is possible only at 100 Hz, though if we focus on the peak of the distribution (50\%) interval, a reasonable area is found when the signal reaches 70 Hz.}
    \label{fig:GW170817_like}
\end{figure*}

\section{Conclusion and Outlook}
\label{sec:concl}

This work presents an ML framework for the detection and characterization of inspiraling BNS signals. We use a frequency domain representation of the data with signals going from a minimum frequency of 20 Hz to a certain maximum frequency which varies between networks. We obtain a good detection rate of inspiraling signals, with better results as the maximum frequency increases. Our neural posterior estimation framework produces a posterior distribution for the chirp mass, the mass ratio, the sky angles, the luminosity distance, and the inclination. These parameters are the most relevant for EM follow-up and, in the future, our network could easily be modified to infer all the binary's parameters. The maximum frequencies chosen in this work are still relatively high since the lowest frequency considered is 40 Hz. This choice is made to ensure enough signal is present in the data to extract an informative sky map at moderate SNR for the chosen maximum frequency. It will be lowered in future work.

First, for the detection part of our framework, we demonstrate its strong performance by analyzing its true alarm probability as a function of the PISNR. We find detection possibilities are comparable to those already presented in the literature for other ML approaches. Unlike these approaches, our method does not rely on matched filtering, meaning we have a reduced alert latency. Moreover, our detection framework is naturally integrated with our neural posterior estimator, as both use the same input data. We also show that our performance remains relatively stable across various false-alarm probabilities, allowing us to decrease the false-alarm rate without significantly reducing the detection rate of genuine triggers.

To demonstrate that our inference algorithm is a viable method for early alert information production, we show it is unbiased by providing a PP-plot, with all curves lying along the diagonal, for all parameters and maximum cut-off frequencies. We also examine several sky maps, showing that our network can make reasonable sky maps when the PISNR of the signal present in the data is large enough. However, for more commonly detected signals, the PISNR for a low maximum frequency is often too small to produce a good sky map. Still, in that case, it is possible to obtain a reasonable estimation of the other parameters relevant to EM follow-up. In particular, in some cases, based on the luminosity distance and inclination estimates, one can already gauge whether the EM counterpart is likely to be observable or not, depending on the expected orientation of the jet. We also analyze the sky map area distribution of our model, confirming that a fraction of the signals have an informative sky map, i.e. small enough to be followed up by electromagnetic means. This proportion increases as more of the signal is observed. This is the expected behavior, and our results are in line with others presented in the literature~\cite{Sachdev:2020lfd, Chatterjee:2022dik}, although we include the estimation of other parameters and do not use information coming from matched filtering pipeline, effectively reducing the latency and potentially increasing accuracy. 

To demonstrate the potential our framework for detection and characterization, we analyze a set of 90 BNS signals, corresponding to five years of observations for a LIGO-Virgo network at design sensitivity. From those, about a third can be detected before their merger, and a bit less than a fifth can be detected at a 40 Hz cut frequency. This is comparable, if not better, than what has been presented in previous works. For the detected events, we also examine their sky maps, showing that they follow an expected trend, with more accurate results for louder signals and for higher maximum frequencies. The best sky maps found for these signals would allow for electromagnetic follow-up.

Finally, we analyzed the detection and inference of realistic signals, taking GW170817 and GW190425 as model signals. We scaled GW170817 to have the same SNR as when the signal was observed in O2 and kept the parameters constant for GW190425. Both signals can be detected before their merger. Moreover, for both we produce a sky map usable for EM follow-up at 100 Hz. For lower cut-off frequencies, sky maps can be reasonable if we focus on the peak probability. Furthermore, the other parameters relevant for electromagnetic follow-up are already informative when the signal reaches 70 Hz.  Still, useful information could be provided at very low-latency to astronomers with this system.

This study focuses on current detectors with a lower frequency cut-off of 20 Hz. This is already a disadvantage for any early-alert system since the signal spends more time in the low-frequency regime. Therefore, better and earlier information would be obtained if we considered an upcoming sensitivity or generation of detectors. In principle, our framework can easily be adapted to incorporate 2G detectors with better sensitivity. For next-generation detectors, however, it would be critical to account for the Earth rotation when training the network since the transition from (1Hz) 5 Hz to 40 Hz will already take about (20) 1.5 hours. This effect enhances our capacity to localize a signal since modulations with Earth's rotation provide information about the event's position. We leave the explicit demonstration for future work.

Another interesting avenue for development is to create a single network capable of dealing with all the maximum frequencies at once, i.e. producing a trigger and posteriors for any signal regardless of time to merger, ideally corresponding to the full signal. The latter is complicated to obtain because of the duration of BNS signals and the required accuracy~\cite{Kolmus:2024scm}. Moreover, this would significantly increase the training cost.
However, it is reasonable to assume we could provide posterior parameter distributions for signals up so a certain maximum frequency without needing the full signal. Those results could then be used as seeds to infer the parameters of the entire signal. This could reduce the computational burden of analyzing the complete data stretch at once without a well-focused prior. 

In conclusion, this work presents a machine learning-based approach to detect and infer parameters relevant for electromagnetic follow-up of inspiraling binary neutron star signals in their pre-merger phase. Our setup does not require input related to matched filtering analyses, reducing the latency. We obtain a good detection rate and accurate posteriors in our analyses. While further work is needed to make this system fully operational, it represents an interesting avenue for future pre-merger and rapid analysis of inspiraling signals with possible electromagnetic counterparts.

\section*{Acknowledgments}
The authors would like to thank Jurriaan Langendorff for interesting discussions on related topics. W.v.S., J.J., and C.V.D.B are supported by the research program of the Netherlands Organisation for Scientific Research (NWO). A.K. and C.V.D.B. are supported by the NWO under the CORTEX project (NWA.1160.18.316). This material is based upon work supported by NSF's LIGO Laboratory which is a major facility fully funded by the National Science Foundation.

\bibliography{bibbestand}

\appendix
\onecolumngrid
\section{Results for the GW190425-like Injection}
\label{app:results_GW190425_like}

In this section, we give the corner plot for the GW190425-like injection, where we do not rescale the SNR. 

\begin{figure*}
    \centering
    \includegraphics[keepaspectratio, width=0.9\textwidth]{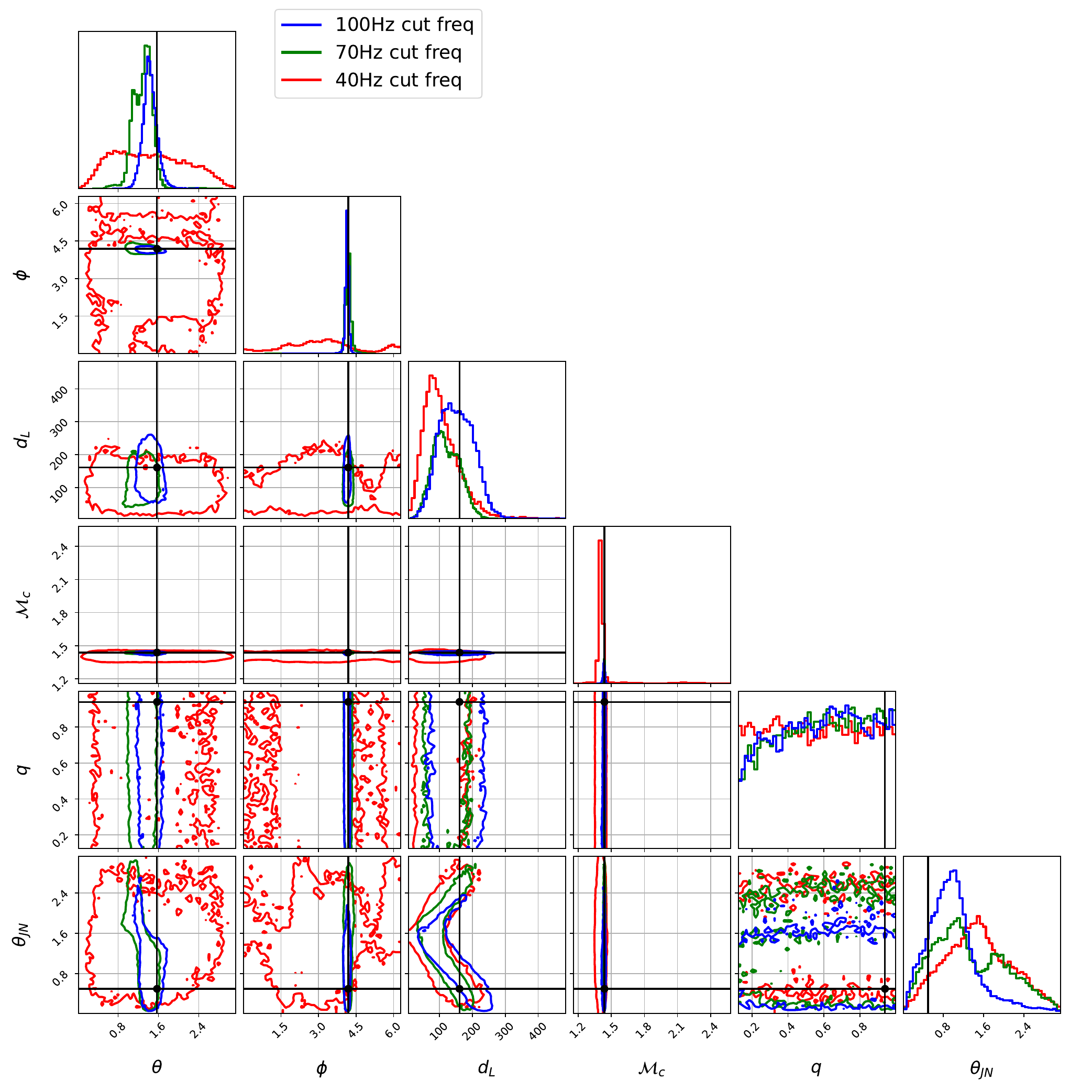}
    \caption{Corner plot for the GW190425-like injection. In this case, the sky map is really poor only for the 50 Hz cut frequency. The two others would already enable one to do a decent follow-up. For the other parameters, the chirp mass is well characterized in all cases, and all of the parameters present an improvement as more of the signal is detected.}
    \label{fig:GW190425_like}
\end{figure*}

\end{document}